\documentclass[11pt,a4paper]{article}

\usepackage{jheppub}
\usepackage{amsmath}

\allowdisplaybreaks[2]

\newcommand{\half}{{\textstyle\frac{1}{2}}}

\newcommand{\ms}{\mskip 1.5mu}
\newcommand{\bs}{\mskip -1.5mu}

\newcommand{\vek}[1]{\boldsymbol{#1}}

\newcommand{\Kp}[1]{K{\smash{'}}_{\!\!\bs #1}}

\subheader{\hfill DESY 12-176}

\title{Angular correlations in the double Drell-Yan process}

\author{Tomas Kasemets}
\author{and Markus Diehl}
\affiliation{Deutsches Elektronen-Synchroton DESY, 22603 Hamburg,
  Germany} 
\emailAdd{tomas.kasemets@desy.de}
\emailAdd{markus.diehl@desy.de}

\abstract{We study the impact of parton correlations on the double
  Drell-Yan process, i.e.\ on the production of two electroweak gauge
  bosons by double parton scattering in a single proton-proton collision.
  Spin correlations between two partons in a proton are shown to change
  the overall rate of the process and to induce characteristic angular
  correlations between the decay leptons of the two gauge bosons.}

\begin{document}

\maketitle

\section{Introduction}

In the era of the LHC, a thorough understanding of the strong-interaction
dynamics in proton-proton collisions is an imperative.  Whereas the
determination of parton distribution functions and the computation of
hard-scattering processes at parton level are becoming fields of precision
physics, other aspects remain much less well understood, both at a
conceptual and a practical level.  Among these aspects are multiparton
interactions, where in a single proton-proton collision more than one
parton in each proton participates in a hard scattering.  Multiparton
interactions can give sizeable contributions both to signal and to
background processes, notably to Higgs production \cite{DelFabbro:1999tf},
to electroweak processes
\cite{Godbole:1989ti,Eboli:1997sv,Cattaruzza:2005nu,Maina:2009vx,%
  Maina:2009sj,Maina:2010vh,Kulesza:1999zh,Gaunt:2010pi,Berger:2011ep} and
to multijet production
\cite{Humpert:1983pw,Humpert:1984ay,Ametller:1985tp,Mangano:1988sq,%
  Drees:1996rw,DelFabbro:2002pw,Domdey:2009bg,Berger:2009cm}.  They are
also of interest in their own right from the point of view of hadron
structure, because they contain information about correlations between
partons inside the proton.

Experimental evidence for hard multiparton interactions has been found at
the ISR \cite{Akesson:1986iv}, the SPS \cite{Alitti:1991rd} and the
Tevatron
\cite{Abe:1993rv,Abe:1997bp,Abe:1997xk,Abazov:2009gc,Abazov:2011rd}.  Due
to the rapid increase of parton densities at small momentum fractions, one
expects these interactions to be even more prominent at the LHC, and first
results have indeed been reported \cite{Dobson:2011co}, with more studies
being expected in the future \cite{Bartalini:2011xj}.  Substantial effort
has gone into the modeling and implementation of multiparton interactions
in Monte-Carlo generators \cite{Sjostrand:1986ep,Sjostrand:1987su,
  Butterworth:1996zw,Sjostrand:2004pf,Corke:2009tk,Corke:2011yy,Bahr:2008dy}.
A mini-review on earlier developments of the subject is given in
\cite{Sjostrand:2004pf}, and an overview of current developments can be
found in the conference proceedings
\cite{Bartalini:2010su,Bartalini:2011jp}.  Recently there has been renewed
interest in understanding the theoretical foundations of multiple hard
scattering
\cite{Blok:2010ge,Blok:2011bu,Diehl:2011tt,Diehl:2011yj,Bartels:2011qi,%
  Ryskin:2011kk,Ryskin:2012qx,Gaunt:2011xd,Gaunt:2012dd,Manohar:2012jr,%
  Manohar:2012pe}, with many issues remaining to be clarified or worked
out.

The simplest assumption one can make when computing multiple hard
scattering is that the different partons in the proton are uncorrelated
with each other.  This leads to very compact results, but it certainly is
a simplification whose validity and limitations need to be investigated.
Recent phenomenological studies of different types of correlation effects
in multiparton interactions can be found in
\cite{Calucci:1997ii,Calucci:1999yz,DelFabbro:2000ds,Rogers:2009ke,%
  Domdey:2009bg,Flensburg:2011kj,Corke:2011yy}.  In this work we focus on
correlations between the \emph{polarization} of two partons in an
unpolarized proton and on their consequences for the overall rate and for
final-state distributions in double hard scattering processes.  We build
on the observations made in \cite{Diehl:2011yj} and extend the results of
\cite{Manohar:2012jr}.  The relevance of spin correlations for multiple
interactions was pointed out long ago in \cite{Mekhfi:1985dv}, but that
work have not been followed up until recently.

We perform our study for double Drell-Yan production, where two
electroweak gauge bosons ($\gamma^*, Z, W^\pm$) are produced in two
independent quark-antiquark annihilation processes.  This has long been
recognized as a prototype for multi-parton interactions
\cite{Goebel:1979mi,Mekhfi:1983az,Halzen:1986ue}.  Unfortunately, recent
phenomenological estimates \cite{Kulesza:1999zh,Gaunt:2010pi,Kom:2011nu}
find that rates for double Drell-Yan production at the LHC will likely be
too small to allow for a detailed experimental investigation of
final-state distributions.  We nevertheless choose this process for our
investigation, given that it is among the simplest double scattering
process from the theoretical point of view but still exhibits a wealth of
nontrivial features.  The results we obtain for double Drell-Yan
production will have their analogs for processes with higher rates, such
as the production of four jets or the production of one gauge boson plus a
dijet.  (Note that up to a global factor the tree-level graph for the
production and subsequent decay of a gauge boson is identical to the graph
for $q_1\ms \bar{q}_2 \to q_3\ms \bar{q}_4$ with a gluon in the
$s$-channel.)

Even if two partons in a proton are unpolarized, there can be correlations
between their transverse positions and their longitudinal momentum
fractions.  There are indeed reasons to expect such correlations, see
e.g.~\cite{Corke:2011yy} and section 2.6 of \cite{Diehl:2011yj}, but not
much is currently known about them.  In the spirit of an exploratory
study, we will calculate these correlations in a simple model of the
proton and discuss their consequence for the transverse-momentum spectrum
of double Drell-Yan production.

This paper is organized as follows.  In the next section we recall the
basic formalism for computing double parton interactions, define double
parton distributions that describe spin correlations between the two
partons and give the polarization dependent parton-level cross sections.
In section~\ref{sec:ddy-cross-sect} we give our results for the cross
section of the double Drell-Yan process.  In section~\ref{sec:model} we
investigate the correlation between transverse and longitudinal variables
in double parton distributions within a simple model.
Section~\ref{sec:summary} summarizes our findings, and in an appendix we
list the coupling factors entering the cross section formulae in
section~\ref{sec:ddy-cross-sect}.


\section{Double parton scattering}
\label{sec:dps}

Consider the production of two gauge bosons $V_1, V_2 = \gamma^*, Z, W$ in
a $pp$ collision, followed by the leptonic decays $\gamma^*, Z \to \ell^+
\ell^-$ or $W \to \ell \nu$.  Four-momenta are assigned as $p(p) +
p(\bar{p}) \to V_1(q_1) + V_2(q_2) + X$.  We are interested in the fully
differential cross section of the four-lepton final state and restrict
ourselves to the kinematic region where the transverse momenta $\vek{q}_1$
and $\vek{q}_2$ of the gauge bosons in the $pp$ center-of-mass are much
smaller than their invariant masses, i.e.\ we assume $\vek{q}_1^2,
\vek{q}_2^2 \ll q_1^2, q_2^2$.  It is in this kinematics that double
parton scattering is not power suppressed compared with the production of
the gauge boson pair by a single hard scattering \cite{Diehl:2011tt}.  In
the calculation of the cross section, the invariant masses $Q_i =
(q_i^2)^{1/2}$ will serve as the hard scale necessary for the application
of factorization.  For simplicity we shall not assume any particular
hierarchy in size between $\vek{q}_1$ and $\vek{q}_2$ or between $Q_1$
and~$Q_2$.

\begin{figure}
\centering
\includegraphics[width=0.65\textwidth]{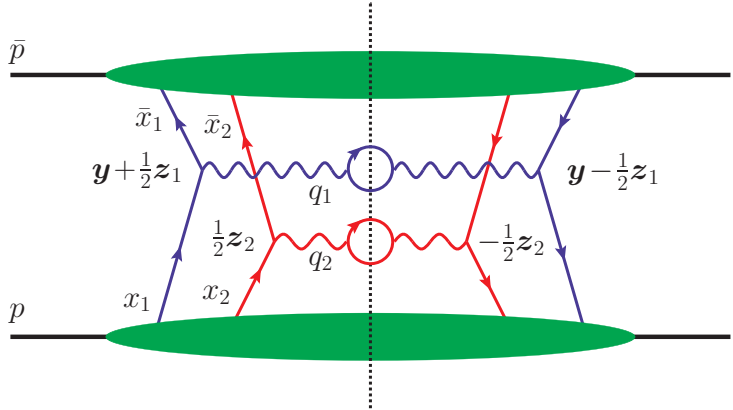}
\caption{\label{fig:dy-graph} A graph for the double Drell-Yan process,
  where two quarks in the right-moving proton interact with two antiquarks
  in the left-moving proton.  The figure shows the assignment of
  four-momenta ($p$, $\bar{p}$, $q_1$, $q_2$), of light-cone momentum
  fractions ($x_i$, $\bar{x}_i$) and of transverse position arguments as
  explained in the text.  The dotted vertical line denotes the final-state
  cut.} 
\end{figure}

We assume that the double hard scattering cross section factorizes into
the product of a double parton distribution (DPD) in each proton and a
parton-level cross section for each of the two hard scatters.  This
factorization has not been proven, but several elements of such a proof
have been given in \cite{Diehl:2011yj}.  Schematically, the double parton
scattering cross section then reads \cite{Diehl:2011tt,Diehl:2011yj}
\begin{align}
\label{eq:cross-simple}
\frac{d\sigma}{\prod_{i=1}^2d x_i\, d\bar{x}_i\, d^2\vek{q}{}_i\,
  d\Omega_i} \,\bigg|_{\text{DPS}}
&= \frac{1}{C}\,
\frac{d\hat{\sigma}_{1}}{d\Omega_1}\,
\frac{d\hat{\sigma}_{2}}{d\Omega_2}
\int\frac{d^2\vek{z}_1}{(2\pi)^2}\, \frac{d^2\vek{z}_2}{(2\pi)^2}\;
e^{-i \vek{z}_1 \vek{q}{}_1 -i \vek{z}_2 \vek{q}{}_2}
\nonumber \\
&\quad \times\int d^2\vek{y}\;
 F(x_1,x_2,\vek{z}_1,\vek{z}_2,\vek{y})\,
 \bar{F}(\bar{x}_1,\bar{x}_2,\vek{z}_1,\vek{z}_2,\vek{y})
\end{align}
with a combinatorial factor $C$ equal to 2 when the final states of the
two hard interactions are identical and equal to 1 otherwise.  Here
$d\hat{\sigma}_i /d\Omega_i$ denotes the cross section for quark-antiquark
annihilation into a lepton pair via the gauge boson $V_i$, taken
differential w.r.t.\ the lepton angles in the appropriate boson rest frame
(see section~\ref{sec:boson-decay-frames}).  In the $pp$ center-of-mass we
define the $z$ axis to point into the direction of the proton momentum $p$
and use light-cone coordinates $v^{\pm} = (v^0\pm v^3)/\sqrt{2}$ and
$\vek{v}=(v^1,v^2)$ for any four-vector $v$.  The kinematic variables
$x_i^{} = q_i^{\smash{+}} /p^+$ and $\bar{x}_i^{} = q_i^{\smash{-}} /
\bar{p}^{\,-}$ determine the longitudinal parton momentum fractions in the
DPDs, which we denote by $F$ for the proton with momentum $p$ and by
$\bar{F}$ for the proton with momentum $\bar{p}$.  The arguments
$\vek{z}_i$ and $\vek{y}$ of the distributions determine where the
hard-scattering processes take place in transverse configuration space.
As indicated in figure~\ref{fig:dy-graph}, $\vek{y}$ is the transverse
distance between the two scattering partons in a proton (and hence between
the two annihilation processes) if one takes the average between the
scattering amplitude and its conjugate.  As shown in
\cite{Diehl:2011tt,Diehl:2011yj} $\vek{z}_i$ is the Fourier conjugate
variable to the transverse momentum of parton $i$ (again averaged between
amplitude and conjugate amplitude).
$F(x_1,x_2,\vek{z}_1,\vek{z}_2,\vek{y})$ is thus the Fourier transform of
a transverse-momentum dependent DPD.  The factorization formula
\eqref{eq:cross-simple} generalizes the expression for single Drell-Yan
production in terms of transverse-momentum dependent single-parton
densities \cite{Collins:1981uk,Collins:2011}.

Integrating \eqref{eq:cross-simple} over the transverse boson momenta
$\vek{q}_1$ and $\vek{q}_2$ one obtains
\begin{align}
\label{eq:cross-coll-simple}
\frac{d\sigma}{\prod_{i=1}^2 d x_i\, d\bar{x}_i\,
  d\Omega_i} \,\bigg|_{\text{DPS}}
&= \frac{1}{C}\,
\frac{d\hat{\sigma}_{1}}{d\Omega_1}\,
\frac{d\hat{\sigma}_{2}}{d\Omega_2}
\int d^2\vek{y}\; F(x_1,x_2,\vek{y})\,
 \bar{F}(\bar{x}_1,\bar{x}_2,\vek{y}) \,.
\end{align}
Here $F(x_1,x_2,\vek{y})$ and $\bar{F}(\bar{x}_1,\bar{x}_2,\vek{y})$ are
transverse-momentum integrated (also called col\-linear) DPDs, which were
introduced long ago in \cite{Paver:1982yp,Mekhfi:1983az}.  Naively, they
are obtained by setting $\vek{z}_1 = \vek{z}_2 = \vek{0}$ in the
distributions that appear in \eqref{eq:cross-simple}.  Closer analysis
reveals that transverse-momentum dependent and collinear DPDs (just as
their counterparts for single partons) require different types of
regularization and subtractions of divergences.  As a result the
distributions depend in different ways on an ultraviolet renormalization
scale, and (with the exception of specific distributions) also on
a rapidity parameter which is closely related to Sudakov logarithms.
This is discussed in \cite{Diehl:2011yj} and will be tacitly implied in
the remainder of the present work.

Equations \eqref{eq:cross-simple} and \eqref{eq:cross-coll-simple} are
schematic in that they omit labels for and summation over the quantum
numbers of the partons (quarks vs.\ antiquarks, flavor, polarization and
color).  This information will be restored in section~\ref{sec:hard-proc}.
We emphasize that these equations only give one contribution to the cross
section for four-lepton production.  Further contributions need to be
added from the familiar single hard-scattering mechanism (where the four
leptons are produced in a single parton-level process), the interference
between single and double hard scattering, as well as double
hard-scattering graphs with fermion number interference
\cite{Diehl:2011yj}.  The single hard-scattering contribution is
straightforward to compute (see e.g.\ \cite{Kom:2011nu}), whereas the
different interference contributions are not.  As argued in
\cite{Diehl:2011yj}, the fermion number interference contribution should
become relatively unimportant at small momentum fractions $x_i$,
$\bar{x}_i$.


\subsection{Double parton distributions}
\label{sec:dpds}

We now briefly specify the DPDs necessary for our study, referring to
\cite{Diehl:2011yj} for more detail.  For two quarks in an unpolarized
right-moving proton we write
\begin{align}
\label{eq:dpds}
F_{a_1a_2}(x_1,x_2,\vek{z}_1,\vek{z}_2,\vek{y})
 &= 2p^+ \!\! \int \frac{dz^-_1}{2\pi}\, \frac{dz^-_2}{2\pi}\, dy^-\;
         e^{i x_1^{} z_1^- p^+ + i x_2^{} z_2^- p^+}
\left<p|\, \mathcal{O}_{a_2}(0,z_2)\, 
           \mathcal{O}_{a_1}(y,z_1) \,|p\right> \,,
\end{align}
where an average over the proton polarization is implied.  The operators
\begin{align}
\label{eq:parton-ops}
\mathcal{O}_{a_i}(y,z_i)
 &= \bar{q}_i(y - \half z_i)\,
     \Gamma_{a_i} q_i(y + \half z_i)\ms \Big|_{z_i^+ = y^+_{\phantom{i}} = 0}
\end{align}
for quarks of flavor $q_i$ are understood to include appropriate Wilson
lines between the two quark fields, as well as appropriate regularization
and subtractions as mentioned above.  Notice the correspondence of
position arguments in \eqref{eq:dpds} and \eqref{eq:parton-ops} with
figure \ref{fig:dy-graph}.  The Dirac matrices
\begin{align}
\Gamma_q &= \half \gamma^+ \,, &
\Gamma_{\Delta q} &= \half \gamma^+\gamma_5 \,, &
\Gamma_{\delta q}^j &= \half i \sigma^{j+} \gamma_5 \quad (j=1,2)
\end{align}
select different polarizations of the quarks in the proton, with $q$
corresponding to unpolarized quarks, $\Delta q$ corresponding to
longitudinal polarization and $\delta q$ to polarization in the transverse
direction $j$.  Notice that the labels $a_i$ in \eqref{eq:dpds} specify
both the flavor and the polarization of the quarks.  In full analogy one
can define DPDs $F_{\bar{a}_1, \bar{a}_2}$ for two antiquarks, as well as
quark-antiquark distributions $F_{\bar{a}_1, a_2}$ and $F_{a_1,
  \bar{a}_2}$.

The quark coupling to gauge boson $V_1$ need not have the same flavor in
the scattering amplitude and its complex conjugate, because a mismatch in
flavor can be compensated by the quark coupling to gauge boson $V_2$.  For
example, the quarks with transverse positions $\vek{y} + \half\vek{z}_1$
and $-\half\vek{z}_2$ in figure~\ref{fig:dy-graph} can be $u$ quarks if
the quarks with transverse positions $\half\vek{z}_2$ and $\vek{y} -
\half\vek{z}_1$ are $d$ quarks.  The DPDs describing this type of quark
flavor interference are given by
\begin{align}
  \label{eq:inter-dpds}
F_{a_1a_2}^I(x_1,x_2,\vek{z}_1,\vek{z}_2,\vek{y})
&= 2p^+ \!\! \int \frac{dz^-_1}{2\pi}\, \frac{dz^-_2}{2\pi}\, dy^-\;
         e^{i x_1^{} z_1^- p^+ + i x_2^{} z_2^- p^+}
  \left<p|\, \mathcal{O}_{a_2}^I(0,z_2)\, 
             \mathcal{O}_{a_1}^I(y,z_1) \,|p\right>
\end{align}
with the product of operators
\begin{align}
  \label{eq:inter-ops}
\mathcal{O}_{a_2}^I(0,z_2)\, \mathcal{O}_{a_1}^I(y,z_1)
&= \bar{q}_1(-\half z_2)\, \Gamma_{a_2}\ms q_2(\half z_2)\;
   \bar{q}_2(y-\half z_1)\, \Gamma_{a_1}\ms q_1(y+\half z_1)
   \Big|_{z_1^+ = z_2^+ = y^+_{\phantom{i}} = 0} \,.
\end{align}
We note that these distributions are complex valued and that their
imaginary part changes sign when one interchanges the flavor (but not the
spin) assignments and replaces $\vek{z}_i \to -\vek{z}_i$, e.g.
\begin{align}
  \label{eq:inter-conjug}
F_{q_1 \Delta q_2}^I(x_1,x_2,\vek{z}_1,\vek{z}_2,\vek{y})
 = \bigl[ F_{q_2 \Delta q_1}^I(x_1,x_2,-\vek{z}_1,-\vek{z}_2,\vek{y}) 
   \bigr]^* \,.
\end{align}
As we shall see, this ensures that physical cross sections are
real-valued.

Let us now classify the different combinations of quark polarization,
taking into account the constraints of parity invariance
\cite{Diehl:2011yj}.  For unpolarized and longitudinally polarized quarks
we have
\begin{align}
\label{eq:DPD1}
F_{qq} &= f_{qq}(x_1,x_2,\vek{z}_1,\vek{z}_2,\vek{y}) \,,
&
F_{\Delta q \Delta q} &=
  f_{\Delta q \Delta q}(x_1,x_2,\vek{z}_1,\vek{z}_2,\vek{y}) \,,
\nonumber \\
F_{q \Delta q} &= g_{q \Delta q} (x_1,x_2,\vek{z}_1,\vek{z}_2,\vek{y}) \,,
&
F_{\Delta q q} &= g_{\Delta q q}(x_1,x_2,\vek{z}_1,\vek{z}_2,\vek{y}) \,, 
\end{align}
where $f$ denotes scalar and $g$ pseudoscalar functions.  For transverse
quark polarization the parton distributions carry a transverse index and
can be decomposed as
\begin{align}
\label{eq:DPD2}
F_{\Delta q \delta q}^i &= M \left(\vek{y}^i f_{\Delta q \delta q}
  + \tilde{\vek{y}}^i g_{\Delta q \delta q}\right) \,,
&
F_{ \delta q\Delta q}^i &= M \left(\vek{y}^i f_{ \delta q\Delta q}
  + \tilde{\vek{y}}^i g_{ \delta q\Delta q}\right) \,,
\nonumber \\
F_{q \delta q}^i  &= M \left(\tilde{\vek{y}}^i f_{q\delta q} 
   + \vek{y}^i g_{q\delta q}\right) \,,
&
F_{\delta q q}^i  &= M \left(\tilde{\vek{y}}^i f_{\delta q q}
   + \vek{y}^i g_{\delta q q}\right) \,,
\end{align}
where the scalar and pseudoscalar functions depend on the same variables
as in \eqref{eq:DPD1}.  Here $\tilde{\vek{y}}^i = \epsilon^{ij} \vek{y}^j$
is a transverse vector orthogonal to $\vek{y}^i$, defined in terms of the
two-dimensional antisymmetric tensor $\epsilon^{ij}$ (with
$\epsilon^{12}=1$).  Factors of the proton mass $M$ have been introduced
in order to have the same mass dimension for all distributions $f$ and
$g$.  For two transversely polarized quarks we finally write
\begin{align}
\label{eq:DPD3}
F_{\delta q \delta q}^{ij} 
 &= \delta^{ij} f_{\delta q \delta q}^{}
   + M^2 \left( 2 \vek{y}^i \vek{y}^j
          - \vek{y}^2 \delta^{ij} \right) f_{\delta q\delta q}^t
\nonumber \\
 &\quad 
  + M^2 \left(\vek{y}^i \tilde{\vek{y}}^j
    + \tilde{\vek{y}}^ i\vek{y}^j\right) g_{\delta q\delta q}^s 
  + M^2 \left(\vek{y}^i \tilde{\vek{y}}^j
    - \tilde{\vek{y}}^ i\vek{y}^j\right) g_{\delta q\delta q}^a \,.
\end{align}
Decompositions analogous to \eqref{eq:DPD1} to \eqref{eq:DPD3} hold for
antiquarks and for flavor interference distributions.

Corresponding definitions apply for two partons in a left-moving proton,
with $+$ and~$-$ components interchanged in \eqref{eq:dpds} to
\eqref{eq:inter-ops}.  Note that the covariant expression of the
two-dimensional antisymmetric tensor in terms of the four-dimensional one
is $\epsilon^{ij} = \epsilon^{+-ij}$ (with $\epsilon_{0123} = 1$).  In the
analogs of \eqref{eq:DPD1} to \eqref{eq:DPD3} for left-moving partons one
hence needs to change the sign of $\tilde{\vek{y}}$ and of the
pseudoscalar functions $g$ (which can be written as $\epsilon^{ij}$
contracted with a parity even tensor constructed from $\vek{z}_1$,
$\vek{z}_2$ and $\vek{y}$).

All distributions discussed so far allow for two color structures, one
where the two fields in the operator $\mathcal{O}_{a_i}$ are coupled to a
color singlet and one where they are coupled to a color octet.  This
requires a further index on all distributions, which we will not display
in the present work for brevity.


\subsection{Reference frames}
\label{sec:boson-decay-frames}

Let us now introduce the reference frames and coordinate axes needed to
describe the angular dependence of the cross section.

In the $pp$ center-of-mass we have the $z$ axis pointing along the
momentum $p$.  The four-vector defining this axis is hence $Z^\mu = (p -
\bar{p})^\mu /\sqrt{2 p\bar{p} \rule{0pt}{1.7ex}}$, where here and in the
following we neglect the proton mass.  We choose a fixed four-vector
$X^\mu$ orthogonal to $p$ and $\bar{p}$ to define the $x$ axis.  The
precise choice does not matter for our purpose, but one may for instance
adopt the convention to have the $x$ direction point towards the center of
the LHC ring.  The $y$ axis is then defined such as to obtain a
right-handed coordinate system; the corresponding four-vector can be
written as $Y^\mu = \epsilon^{\mu}{}_{\nu\rho\sigma}\, X^\nu \bar{p}^\rho
p^\sigma /(p\bar{p})$.

\begin{figure}
\begin{center}
\includegraphics[width=0.66\textwidth]{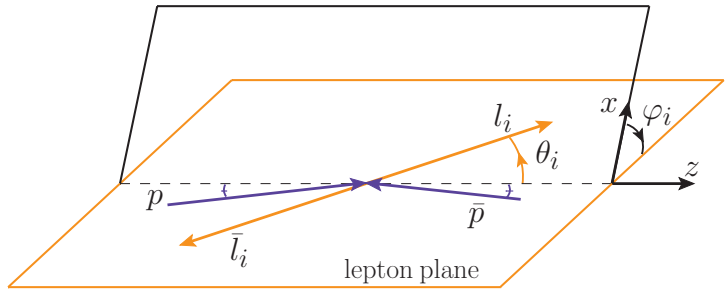}
\end{center}
\caption{\label{figures:angles} Coordinate system in the rest frame of
  vector boson $V_i$.  The $z$ axis bisects the angle between the spatial
  components of the momenta $p$ and $-\bar{p}$, and the $x$ axis
  corresponds to a fixed reference direction as explained in the text.
  (In general, the proton momenta are therefore not in the $x$-$z$ plane.)
  $l_i$ and $\bar{l}_i$ are the momenta of the lepton and the antilepton
  from the boson decay, respectively.  $\theta_i$ denotes the polar and
  $\varphi_i$ the azimuthal angle of the lepton.  Note that $\varphi_i$ is
  negative in this example.}
\end{figure}

The kinematics of the gauge boson decays into lepton pairs is conveniently
described in the rest frame of the respective gauge boson.  The $z$ axis
in the rest frame of the boson $V_i$ is defined by the four-vector
\begin{align}
  \label{eq:z-axis}
Z_i^\mu &= \frac{1}{2} \sqrt{Q_i^2 + \vek{q}_i^2}\;
  \biggl[ \frac{p^\mu}{p\ms q_i}
        - \frac{\bar{p}^\mu}{\bar{p}\ms q_i} \biggr] \,,
\end{align}
where $\vek{q}_i$ is the transverse boson momentum in the $pp$
center-of-mass as before.  As illustrated for one boson in
figure~\ref{figures:angles}, the $z$ axis bisects the angle between the
spatial components of $p$ and $-\bar{p}$ in the boson rest frame.  The $x$
axis is specified by
\begin{align}
  \label{eq:x-axis}
X_i^\mu &=
    \frac{1}{\sqrt{1 + (X q_i^{})^2_{\phantom{i}} /Q_i^2}}\,
    \biggl[ X^\mu - \frac{X q_i}{Q_i^2}\, q_i^\mu \biggr]
\end{align}
and the $y$ axis is again defined so as to obtain a right-handed
coordinate system, i.e.\ by $Y_i^\mu = \epsilon^{\mu}{}_{\nu\rho\sigma}\,
Z_i^\nu\, X_{\ms i}^{\smash{\rho}}\; q_i^\sigma /Q_i^{}$.
With these reference axes we define the polar and azimuthal angles
$\theta_i$ and $\varphi_i$ of the lepton (as opposed to the antilepton) in
the decay of $V_i$, i.e.\ of $\ell^-$ in the decay of a $\gamma^*$, $Z$ or
$W^-$ and of $\nu_\ell$ in the decay of a $W^+$.

Noting that $X q_i$ is the $x$ component of $q_i$ in the $pp$
center-of-mass, we see in \eqref{eq:x-axis} that $X_1$, $X_2$ and $X$
differ from each other by amounts of order $|\vek{q}_i| /Q_i^{}$, which is
a small parameter in our calculation of the cross section.  Likewise, one
finds differences of order $|\vek{q}_i| /Q_i^{}$ between $Y_1$, $Y_2$ and
$Y$.  As we shall see shortly, this greatly simplifies the discussion of
azimuthal angles in our calculation.

Readers familiar with the analysis of single Drell-Yan production will
recognize that our choice of $z$ axes is the same as in the Collins-Soper
frame \cite{Collins:1977iv}.  Useful information about this frame can
e.g.\ be found in \cite{Boer:2006eq,Arnold:2008kf}.  By contrast, we
define $x$ axes (and thus the azimuthal angles $\varphi_i$) starting from
a fixed direction in space, whereas in the Collins-Soper frame the $x$
axis is defined such that the proton momenta lie in the $x$-$z$ plane.
The latter choice becomes undefined when the transverse boson momentum in
the $pp$ center-of-mass goes to zero.  For unpolarized single Drell-Yan
production this is not a problem because in this limiting case all
azimuthal dependence in the cross section must vanish due to rotation
invariance.  However, in the double Drell-Yan process there can be an
azimuthal dependence even if $\vek{q}_1$ or $\vek{q}_2$ or both go to
zero, as we shall see.  Choosing one of these vectors (or any linear
combination of them) to define $x$ axes would therefore entail ill-defined
azimuthal angles at some point in phase space where there can be a
nontrivial azimuthal dependence.

The physical cross section must of course not depend on the arbitrary
fixed direction specified by $X^\mu$.  To understand how this happens, we
anticipate that our results will depend only on the \emph{difference} of
azimuthal angles whose definition depends on $X^\mu$, such as for instance
$\varphi_1 - \varphi_2$.  These angles are defined in different frames,
but to the accuracy of our calculation we can replace them with the
azimuthal angles of the leptons in the $pp$ center-of-mass.  This is
easily seen by writing trigonometric functions of $\varphi_i$ in terms of
invariant products $X_i\ms l_i$ and $Y_i\ms l_i$, where $l_i$ is the
four-momentum of the lepton from the decay of $V_i$.  When calculating the
cross section we neglect terms of order $|\vek{q}_i| /Q_i^{}$ and can
hence approximate $X_i\ms l_i \approx X l_i$ and $Y_i\ms l_i \approx Y
l_i$, which gives azimuthal angles in the $pp$ center-of-mass as
announced.


\subsection{Hard-scattering cross sections}
\label{sec:hard-proc}

If we restore labels for quarks and antiquarks, their flavor and their
polarization, the cross section formula \eqref{eq:cross-simple} reads
\begin{align}
  \label{eq:cross}
& \frac{d\sigma}{\prod_{i=1}^2
   dx_i\, d\bar{x}_i\, d^2\vek{q}{}_i\, d\Omega_i}
= \frac{1}{C}\, \sum_{a_1 a_2 a_3 a_4}
   \int\frac{d^2\vek{z}_1}{(2\pi)^2}\, \frac{d^2\vek{z}_2}{(2\pi)^2}\;
     e^{-i \vek{z}_1 \vek{q}{}_1 -i \vek{z}_2 \vek{q}{}_2}
   \int d^2\vek{y}
\nonumber \\
& \quad \times
\biggl[
   \frac{d\hat{\sigma}_{a_1 \bar{a}_3}}{d\Omega_1}\,
   \frac{d\hat{\sigma}_{a_2 \bar{a}_4}}{d\Omega_2}\,
     F_{a_1 a_2} \bar{F}_{\bar{a}_3 \bar{a}_4}
 + \frac{d\hat{\sigma}_{a_1 \bar{a}_3}}{d\Omega_1}\,
   \frac{d\hat{\sigma}_{\bar{a}_2 a_4}}{d\Omega_2}\,
     F_{a_1 \bar{a}_2} \bar{F}_{\bar{a}_3 a_4}
\nonumber \\
& \qquad
 + \frac{d\hat{\sigma}_{\bar{a}_1 a_3}}{d\Omega_1}\,
   \frac{d\hat{\sigma}_{a_2 \bar{a}_4}}{d\Omega_2}\,
     F_{\bar{a}_1 a_2} \bar{F}_{a_3 \bar{a}_4}
 + \frac{d\hat{\sigma}_{\bar{a}_1 a_3}}{d\Omega_1}\,
   \frac{d\hat{\sigma}_{\bar{a}_2 a_4}}{d\Omega_2}\,
     F_{\bar{a}_1 \bar{a}_2} \bar{F}_{a_3 a_4}
\biggr]
\nonumber \\
& \quad + \left\{\text{flavor interference}\right\} \,,
  \phantom{\int}
\end{align}
where here and in the following we omit the label ``DPS'' for double
parton scattering.  In all terms the DPDs have arguments as in
\eqref{eq:cross-simple}, which will be omitted henceforth for brevity.  To
distinguish the distributions for the left- and right-moving proton we use
the notation $F$ and $\bar{F}$ and a corresponding notation for the scalar
and pseudoscalar functions $f, \bar{f}$ and $g, \bar{g}$ introduced in
section~\ref{sec:dpds}.  The first subscript in $d\sigma_{a_i \bar{a}_j}$
and $d\sigma_{\bar{a}_i a_j}$ denotes the right-moving parton and the
second subscript the left-moving one.  The sum over $a_1$ to $a_4$ runs
over quark flavors and polarizations ($q,\Delta q, \delta q$).

The flavor interference terms involve the interference DPDs in
\eqref{eq:inter-dpds} and corresponding interference terms for the hard
scattering.  These interference terms only appear if the produced bosons
are both neutral or both charged, otherwise the quark and antiquark
flavors in the annihilation processes do not match.  We will return to
this in the next section.

Labels for the color structure of the DPDs are not displayed in
\eqref{eq:cross}.  With the conventions of \cite{Diehl:2011yj}, each
factor $F\, \bar{F}$ is to be replaced with the sum ${}^{1\!}F\,
{}^{1\!}\bar{F} + {}^{8\!}F\, {}^{8\!}\bar{F}$ of color singlet and color
octet distributions, without change in the hard-scattering cross sections.
This holds for the production of arbitrary color-neutral states in the
hard-scattering processes.

It is straightforward to compute the tree-level cross section for
quark-antiquark annihilation into a gauge boson followed by its leptonic
decay.  In accordance with the power counting scheme underlying the cross
section formula \eqref{eq:cross}, the transverse boson momenta $\vek{q}_i$
are set to zero in this calculation since by assumption they are small
compared with the invariant mass $Q_i$.  This also simplifies the
kinematics of the gauge boson decays as we already noticed in
section~\ref{sec:boson-decay-frames}.

Consider first the case where both quark and antiquark are unpolarized or
longitudinally polarized.  The angular dependence of the cross section is
then of the form
\begin{align}
  \label{eq:long-pol-cross}
\frac{d\hat{\sigma}_{a_i \bar{a}_j}}{d\Omega_i}
&= \bigl( 1+\cos^2\theta_i \bigr) K_{a_i \bar{a}_j}(Q_i)\,
    + 2 \cos\theta_i\, \Kp{a_i \bar{a}_j}(Q_i) \,,
\qquad\qquad
\end{align}
with $a_i = q_i, \Delta q_i$ and $\bar{a}_j = \bar{q}_j, \Delta\bar{q}_j$.
The integration element reads $d\Omega_i = d\varphi_i\, d\!\cos\theta_i$
as usual.  The factors $K$ and $K'$ depend on coupling constants and on
$Q_i$ via the gauge boson propagators.  One easily finds
\begin{align}
  \label{eq:coupl-fact-relations}
K_{\Delta q_i \Delta\bar{q}_j} &= - K_{q_i \bar{q}_j} \,,
&
K_{q_i \Delta\bar{q}_j} &= - K_{\Delta q_i \bar{q}_j}
\end{align}
and analogous relations for $K'$, so that
\begin{align}
\frac{d\hat{\sigma}_{q_i \bar{q}_j}}{d\Omega_i}
 &= - \frac{d\hat{\sigma}_{\Delta q_i \Delta\bar{q}_j}}{d\Omega_i} \,,
&
\frac{d\hat{\sigma}_{q_i \Delta\bar{q}_j}}{d\Omega_i}
 & = - \frac{d\hat{\sigma}_{\Delta q_i \bar{q}_j}}{d\Omega_i} \,.
\end{align}
Because of chirality conservation for massless quarks one has vanishing
parton-level cross sections for the annihilation of a transversely
polarized parton with an unpolarized or longitudinally polarized one,
$d\hat{\sigma}_{\delta q_i \bar{q}_j} =d \hat{\sigma}_{\delta q_i \Delta
  \bar{q}_j} = d\hat{\sigma}_{q_i \delta \bar{q}_j} =
d\hat{\sigma}_{\Delta q_i \delta \bar{q}_j}=0$.  If both quark and
antiquark are transversely polarized, one finds
\begin{align}
  \label{eq:transv-pol-cross}
\frac{d\hat{\sigma}_{\delta q_i \delta\bar{q}_j}^{kl}}{d\Omega_i} =
 \sin^2\theta_i\, \biggl\{ \! & \Bigl[\,
   \cos(2\varphi_i)\,  K_{\delta q_i \delta\bar{q}_j}(Q_i)
 - \sin(2\varphi_i)\, \Kp{\delta q_i \delta\bar{q}_j}(Q_i) \,\Bigr]
\bigl( \vek{X}^k \vek{X}^l - \vek{Y}^k \vek{Y}^l \bigr)
\nonumber \\
 +\, & \Bigl[\,
   \sin(2\varphi_i)\,  K_{\delta q_i \delta\bar{q}_j}(Q_i)
 + \cos(2\varphi_i)\, \Kp{\delta q_i \delta\bar{q}_j}(Q_i) \,\Bigr]
\bigl( \vek{X}^k \vek{Y}^l \!+ \vek{Y}^k \vek{X}^l \bigr) \!\biggr\}
\end{align}
with $X$ and $Y$ as defined in section~\ref{sec:boson-decay-frames}.  The
transverse indices $k,l$ in \eqref{eq:transv-pol-cross} refer to the $pp$
center-of-mass, where they are to be contracted with the corresponding
indices of the DPDs.  We note that contraction of
\eqref{eq:transv-pol-cross} with the transverse spin vectors $\vek{s}^k$,
$\bar{\vek{s}}^l$ of the quark and the antiquark gives the simple
expression
\begin{align}
  \label{eq:transv-pol-simple}
\frac{d\hat{\sigma}_{\delta q_i \delta\bar{q}_j}^{kl}}{d\Omega_i}\,
\vek{s}^k \bar{\vek{s}}^l &= \sin^2\theta_i
\bigl[ \cos(\varphi_s + \varphi_{\bar{s}} - 2 \varphi_i)\,
        K_{\delta q_i \delta\bar{q}_j}
     + \sin(\varphi_s + \varphi_{\bar{s}} - 2 \varphi_i)\,
       \Kp{\delta q_i \delta\bar{q}_j} \bigr] \,,
\end{align}
where $\varphi_s$ and $\varphi_{\bar{s}}$ are the azimuthal angles of the
spin vectors in the $pp$ center-of-mass and our normalization convention
is $\vek{s}^2 = \vek{\bar{s}}^2 = 1$.

The preceding expressions hold for both neutral and charged vector bosons,
and the coupling factors $K_{a_i \bar{a}_j}$ and $\Kp{a_i \bar{a}_j}$
appearing in \eqref{eq:long-pol-cross} and \eqref{eq:transv-pol-cross} are
given in appendix \ref{sec:coupling-factors}.  For neutral boson
production the annihilating quark and antiquark have the same flavor.  In
this case we will use $d\hat{\sigma}_{q_i \bar{q}_j}$, $d\hat{\sigma}_{q_i
  \Delta \bar{q}_j}$, \ldots with $i\neq j$ to denote the interference
terms for flavor $q_i$ in the amplitude and flavor $q_j$ in the conjugate
amplitude.  The relations \eqref{eq:long-pol-cross} to
\eqref{eq:transv-pol-simple} remain valid for these interference terms.
As can be seen in appendix~\ref{sec:coupling-factors}, the corresponding
coupling factors are complex, and their imaginary parts change sign when
the flavor (but not the spin) labels are interchanged, e.g.
\begin{align}
  \label{eq:inter-coupl}
K_{q_1 \bar{q}_2} &= (K_{q_2 \bar{q}_1})^* \,,
&
\Kp{q_1 \Delta\bar{q}_2} &= (\Kp{q_2 \Delta\bar{q}_1})^* \,.
\end{align}
We note that for invariant masses $Q_i$ far below the $Z$ mass, the
neutral boson channel is well approximated by $\gamma^*$ production alone.
The only nonzero coupling factors in this case are $K_{q_i \bar{q}_j} = -
K_{\Delta q_i \Delta\bar{q}_j} = K_{\delta q_i \delta\bar{q}_j}$.

For $W$ boson production we use $d\hat{\sigma}_{q_i \bar{q}_j}$,
$d\hat{\sigma}_{q_i \Delta \bar{q}_j}$ etc.\ to denote cross sections with
different flavors $q_i$, $q_j$ in the initial state.  We do not need a
separate notation for flavor interference terms in this case, because the
product $d\hat{\sigma}_{a_1 \bar{a}_3}\, d\hat{\sigma}_{a_2 \bar{a}_4}$ of
cross sections for $W W$ production is equal to the product of the
corresponding interference terms, except for CKM factors that can easily
be identified.
Using that $W$ bosons only couple to left-handed fermions, we find further
simplifications for the coupling factors:
\begin{align}
  \label{eq:coupl-W-relations}
K_{q_i \bar{q}_j} &= K_{q_i \Delta\bar{q}_j} \,,
&
K_{\delta q_i \delta\bar{q}_j} &= 0 \,,
&
\Kp{a_1 \bar{a}_2} &= K_{a_1 \bar{a}_2} \,,
\end{align}
where the second relation reflects that the operator $\mathcal{O}_{\delta
  q}$ for transverse quark polarization corresponds to the interference
between left- and right-handed quarks.  Together with the relations
\eqref{eq:coupl-fact-relations} we are thus left with only one independent
coupling factor for $W^-$ and only one for $W^+$ production.

So far we have discussed cross sections and interference terms
$d\hat{\sigma}_{a_i \bar{a}_j}$ for the annihilation of a right-moving
quark with a left-moving antiquark.  The cross sections and interference
terms $d\hat{\sigma}_{\bar{a}_j a_i}$ for right-moving antiquarks and
left-moving quarks have the same form as in \eqref{eq:long-pol-cross} and
\eqref{eq:transv-pol-cross}.  The associated coupling factors are given by
\begin{align}
  \label{eq:qbar-coupl-1}
K_{\bar{q}_j q_i}  &=   \bigl(  K_{q_i \bar{q}_j} \bigr)^* \,,
&
\Kp{\bar{q}_j q_i} &= - \bigl( \Kp{q_i \bar{q}_j} \bigr)^*
\intertext{and analogous relations for the spin combinations $\Delta q\ms
  \Delta q$ and $\delta q\ms \delta q$, and by}
  \label{eq:qbar-coupl-2}
K_{\bar{q}_j \Delta q_i}  &= - \bigl( K_{q_i \Delta \bar{q}_j} \bigr)^* \,,
&
\Kp{\bar{q}_j \Delta q_i} &=  \bigl( \Kp{q_i \Delta \bar{q}_j} \bigr)^*
\end{align}
and an analogous relation for the spin combination $\Delta q\ms q$.


\section{The double Drell-Yan cross section}
\label{sec:ddy-cross-sect}

Inserting the hard-scattering cross sections \eqref{eq:long-pol-cross},
\eqref{eq:transv-pol-cross} and the DPD decompositions \eqref{eq:DPD1} to
\eqref{eq:DPD3} into the factorization formula \eqref{eq:cross}, we obtain
our final results for the double parton scattering contribution to
four-lepton production.

\begin{figure}
\begin{center}
\includegraphics[width=0.65\textwidth]{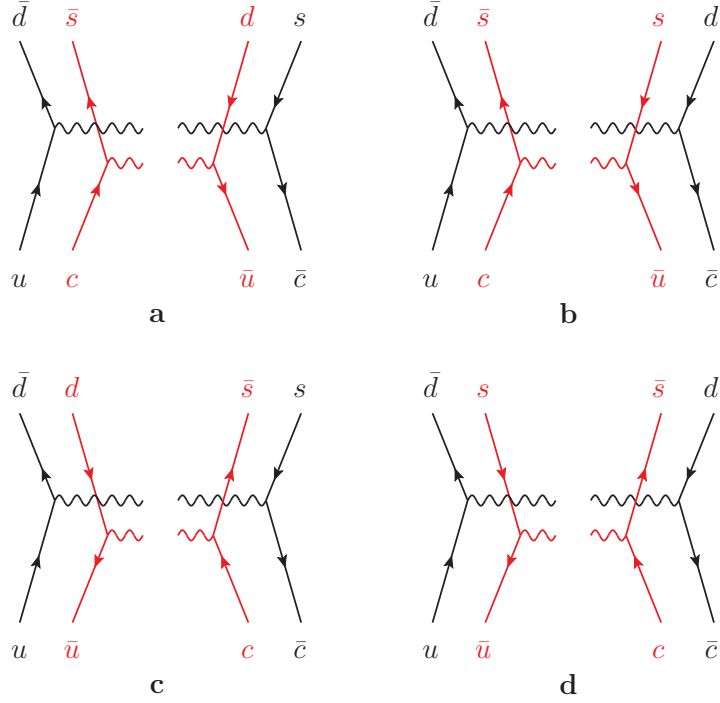}
\end{center}
\caption{\label{figures:interf-W} Hard-scattering graphs for the
  production of $W^+ W^+$ (a, b) or of $W^+ W^-$ (c, d).  The labels $q$
  and $\bar{q}$ indicate whether a parton corresponds to a quark field or
  a conjugate quark field in the relevant DPD.  Graphs (b) and (d) are
  multiplied with interference distributions for one of the protons,
  whereas graphs (a) and (c) go along with interference distributions for
  both protons.}
\end{figure}

For the production and decay of two $W$ bosons, the result has a simple
structure thanks to the relations \eqref{eq:coupl-W-relations},
\begin{align}
  \label{eq:cross-WW}
& \frac{d\sigma^{WW}}{\prod_{i=1}^2
   dx_i\, d\bar{x}_i\, d^2\vek{q}{}_i\, d\Omega_i}
= \frac{1}{C} \! \sum_{q_1 q_2 q_3 q_4} \!
    K_{q_1\bar{q}_3}(Q_1)\, K_{q_2\bar{q}_4}(Q_2)
  \int\frac{d^2\vek{z}_1}{(2\pi)^2}\, \frac{d^2\vek{z}_2}{(2\pi)^2}\;
    e^{-i \vek{z}_1 \vek{q}{}_1 -i \vek{z}_2 \vek{q}{}_2}
  \int d^2\vek{y}
\nonumber \\
& \quad \times \Bigl[
 (1 + \cos\theta_1)^2 \, (1 + \cos\theta_2)^2
\nonumber \\[-0.1em]
& \qquad\quad \times 
 ( f_{q_1 q_2} + f_{\Delta q_1 \Delta q_2}
   - g_{q_1 \Delta q_2} - g_{\Delta q_1 q_2} )
   ( \bar{f}_{\bar{q}_3 \bar{q}_4}
   + \bar{f}_{\Delta \bar{q}_3 \Delta\bar{q}_4}
   - \bar{g}_{\bar{q}_3 \Delta\bar{q}_4}
   - \bar{g}_{\Delta \bar{q}_3 \bar{q}_4} )
\nonumber \\[0.4em]
& \qquad +
  (1 + \cos\theta_1)^2 \, (1 - \cos\theta_2)^2
\nonumber \\[0.1em]
& \qquad\quad \times 
 ( f_{q_1 \bar{q}_4} - f_{\Delta q_1 \Delta \bar{q}_4}
   + g_{q_1 \Delta \bar{q}_4} - g_{\Delta q_1 \bar{q}_4} )
   ( \bar{f}_{\bar{q}_3 {q}_2}
   - \bar{f}_{\Delta \bar{q}_3 \Delta {q}_2}
   + \bar{g}_{\bar{q}_3 \Delta {q}_2}
   - \bar{g}_{\Delta \bar{q}_3 {q}_2} )
\nonumber \\[0.4em]
& \qquad +
  (1 - \cos\theta_1)^2 \, (1 + \cos\theta_2)^2
\nonumber \\[0.1em]
& \qquad\quad \times 
 ( f_{\bar{q}_3 q_2} - f_{\Delta \bar{q}_3 \Delta q_2}
   - g_{\bar{q}_3 \Delta q_2} + g_{\Delta \bar{q}_3 q_2} )
   ( \bar{f}_{{q}_1 \bar{q}_4}
   - \bar{f}_{\Delta {q}_1 \Delta\bar{q}_4}
   - \bar{g}_{{q}_1 \Delta\bar{q}_4}
   + \bar{g}_{\Delta {q}_1 \bar{q}_4} )
\nonumber \\[0.4em]
& \qquad +
  (1 - \cos\theta_1)^2 \, (1 - \cos\theta_2)^2
\nonumber \\[0.1em]
& \qquad\quad \times 
 ( f_{\bar{q}_3 \bar{q}_4} + f_{\Delta \bar{q}_3 \Delta \bar{q}_4}
   + g_{\bar{q}_3 \Delta \bar{q}_4} + g_{\Delta \bar{q}_3 \bar{q}_4} )
   ( \bar{f}_{{q}_1 {q}_2}
   + \bar{f}_{\Delta {q}_1 \Delta {q}_2}
   + \bar{g}_{{q}_1 \Delta {q}_2}
   + \bar{g}_{\Delta {q}_1 {q}_2} )
\nonumber \\[0.1em]
& \qquad
  + \left\{ \text{flavor interference} \right\} \Bigr] \,,
\end{align}
where the sum over $q_1$ to $q_4$ runs over quark flavors.  The flavor
interference terms are obtained by replacing the DPDs in one or in both
protons with their interference analogs and by appropriately changing the
CKM factors in the product $K_{q_1\bar{q}_3}\, K_{q_2\bar{q}_4}$.
Different types of flavor interference terms are shown in
figure~\ref{figures:interf-W}.
Taking into account the minus sign in the definition of pseudoscalar
distributions for left-moving partons, e.g.\ in $\bar{F}_{q_i \Delta
  \bar{q}_j} = - \bar{g}_{q_i \Delta \bar{q}_j}$, we recognize that the
DPD combinations in \eqref{eq:cross-WW} correspond to negative-helicity
quarks and positive-helicity antiquarks, as required by the left-handed
nature of the charged weak current.

We see that for $W$ pair production the presence of longitudinal parton
spin correlations in the proton changes the overall rate of the cross
section as well as the distribution in the polar angles of the decay
leptons.

\begin{figure}
\begin{center}
\includegraphics[width=0.65\textwidth]{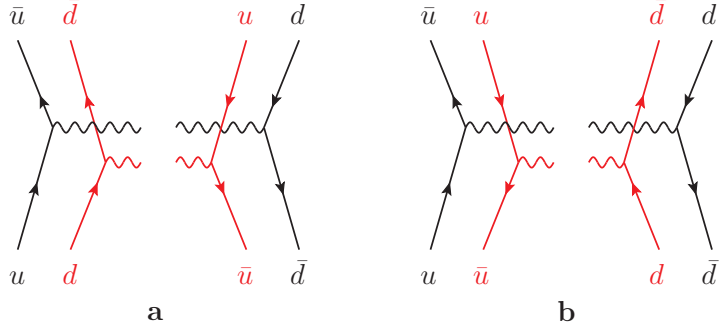}
\end{center}
\caption{\label{figures:interf-Z} Hard-scattering graphs for the
  production of two neutral gauge bosons. The labels $q$ and $\bar{q}$
  have the same meaning as in figure~\protect\ref{figures:interf-W}.}
\end{figure}

For one or two neutral bosons ($\gamma^*, Z$) the structure of the cross
section is more complicated.  We split the cross section \eqref{eq:cross}
into three parts, $\sigma^{(0)}$ for the case without transverse quark
polarization and $\sigma^{(1)}$, $\sigma^{(2)}$ for the cases where one or
two hard interactions are initiated by transversely polarized quarks.  The
contribution with only unpolarized and longitudinally polarized partons
reads
\begin{align}
  \label{eq:zeroT}
& \frac{d\sigma^{(0)}}{\prod_{i=1}^2
   dx_i\, d\bar{x}_i\, d^2\vek{q}{}_i\, d\Omega_i}
= \frac{1}{C}\, 
  \sum_{q_1 q_2 q_3 q_4}
  \int\frac{d^2\vek{z}_1}{(2\pi)^2}\, \frac{d^2\vek{z}_2}{(2\pi)^2}\;
    e^{-i \vek{z}_1 \vek{q}{}_1 -i \vek{z}_2 \vek{q}{}_2}
  \int d^2\vek{y}
\nonumber \\
& \quad \times \Bigl\{
  \left[ (1+\cos^2\theta_1) K_{q_1 \bar{q}_3}
        + 2 \cos\theta_1\, \Kp{q_1 \bar{q}_3} \right]
  \left[ (1+\cos^2\theta_2) K_{q_2 \bar{q}_4}
        + 2 \cos\theta_2\, \Kp{q_2 \bar{q}_4} \right] 
\nonumber \\[-0.1em]
& \qquad\quad \times \left(
   f_{q_1 q_2} \bar{f}_{\bar{q}_3 \bar{q}_4}
 + f_{\Delta q_1\Delta q_2} \bar{f}_{\Delta\bar{q}_3 \Delta \bar{q}_4}
 + g_{q_1 \Delta q_2} \bar{g}_{\bar{q}_3 \Delta\bar{q}_4}
 + g_{\Delta q_1 q_2} \bar{g}_{\Delta \bar{q}_3 \bar{q}_4} \right)
\nonumber \\[0.4em]
& \qquad +
  \left[ (1+\cos^2\theta_1) K_{q_1 \Delta \bar{q}_3}
        + 2 \cos\theta_1\, \Kp{q_1 \Delta \bar{q}_3} \right]
  \left[ (1+\cos^2\theta_2) K_{q_2 \Delta \bar{q}_4}
        + 2 \cos\theta_2\, \Kp{q_2 \Delta \bar{q}_4} \right]
\nonumber \\[0.1em]
& \qquad\quad \times \left(
   f_{q_1 q_2} \bar{f}_{\Delta \bar{q}_3 \Delta \bar{q}_4}
 + f_{\Delta q_1 \Delta q_2} \bar{f}_{\bar{q}_3 \bar{q}_4}
 + g_{q_1 \Delta q_2} \bar{g}_{\Delta \bar{q}_3 \bar{q}_4}
 + g_{\Delta q_1 q_2} \bar{g}_{\bar{q}_3\Delta \bar{q}_4} \right)
\nonumber \\[0.4em]
& \qquad -
  \left[ (1+\cos^2\theta_1) K_{q_1 \bar{q}_3}
        + 2 \cos\theta_1\, \Kp{q_1 \bar{q}_3} \right]
  \left[ (1+\cos^2\theta_2) K_{q_2 \Delta \bar{q}_4}
        + 2 \cos\theta_2\, \Kp{q_2 \Delta \bar{q}_4} \right] 
\nonumber \\[0.1em]
& \qquad\quad \times \left(
   g_{q_1 \Delta q_2} \bar{f}_{\bar{q}_3 \bar{q}_4} 
 + g_{\Delta q_1 q_2} \bar{f}_{\Delta\bar{q}_3 \Delta \bar{q}_4}
 + f_{q_1 q_2} \bar{g}_{\bar{q}_3 \Delta\bar{q}_4}
 + f_{\Delta q_1 \Delta q_2} \bar{g}_{\Delta \bar{q}_3 \bar{q}_4} \right)
\nonumber \\[0.4em]
& \qquad -
  \left[ (1+\cos^2\theta_1) K_{q_1 \Delta \bar{q}_3}
        + 2 \cos\theta_1\, \Kp{q_1 \Delta \bar{q}_3} \right]
  \left[ (1+\cos^2\theta_2) K_{q_2 \bar{q}_4}
        + 2 \cos\theta_2\, \Kp{q_2 \bar{q}_4} \right]
\nonumber \\[0.1em]
& \qquad\quad \times \left(
   g_{q_1 \Delta q_2} \bar{f}_{\Delta \bar{q}_3 \Delta \bar{q}_4} 
 + g_{\Delta q_1 q_2} \bar{f}_{\bar{q}_3 \bar{q}_4}
 + f_{q_1 q_2} \bar{g}_{\Delta \bar{q}_3 \bar{q}_4}
 + f_{\Delta q_1 \Delta q_2} \bar{g}_{\bar{q}_3 \Delta \bar{q}_4} \right)
\Bigr\}
\nonumber \\[0.2em]
& \quad + \{ \text{flavor interference} \}
         + \{ q \bar{q}~\text{permutations} \} \,.
\end{align}
The $q\bar{q}$ permutation terms are obtained by permutation of the
quark-antiquark assignments in the DPDs and in the coupling factors $K$,
$K'$ as specified in \eqref{eq:cross}.  
For neutral bosons the annihilating quark and antiquark have the same
flavor, i.e.\ one has $q_1=q_3$ ($q_2=q_4$) if $V_1$ ($V_2$) is neutral.
The flavor interference term for neutral boson pairs is then obtained by
replacing all distributions $f$, $g$, $\bar{f}$, $\bar{g}$ with their
interference analogs $f^I$, $g^I$, $\bar{f}^I$, $\bar{g}^I$ and by
interchanging $1\leftrightarrow 2$ in the second subscript of the coupling
factors, e.g.\ $K_{q_1 \bar{q}_1} \Kp{q_2 \Delta\bar{q}_2} \to K_{q_1
  \bar{q}_2} \Kp{q_2 \Delta\bar{q}_1}$.  The relations
\eqref{eq:inter-conjug}, \eqref{eq:inter-coupl} and their analogs for
other polarizations ensure that the sum over all flavor assignments in
\eqref{eq:zeroT} gives a real-valued cross section.  Example graphs for
flavor interference are shown in figure~\ref{figures:interf-Z}.

We see in \eqref{eq:zeroT} that longitudinal parton spin correlations
change the overall rate of double parton scattering and the dependence on
the polar angles of the leptons, due to the differences between the
coupling factors $K_{q_i \bar{q}_j}$, $\Kp{q_i \bar{q}_j}$ and $K_{q_i
  \Delta\bar{q}_j}$, $\Kp{q_i \Delta\bar{q}_j}$.  Only in the neutral
boson channel at $Q_i$ values small enough to neglect $Z$ production does
one have a fixed angular dependence $d\sigma^{(0)} /d\cos\theta_i \propto
1+\cos^2\theta_i$.

We now turn towards the part of the cross section where one of the two
annihilation processes involves transverse quark polarization (and thus
produces a neutral gauge boson).  It reads
\begin{align}
  \label{eq:oneT}
& \frac{d\sigma^{(1)}}{\prod_{i=1}^2
   dx_i\, d\bar{x}_i\, d^2\vek{q}{}_i\, d\Omega_i}
= \frac{1}{C}\; \sin^2\theta_2
  \sum_{q_1 q_2 q_3}
  \int\frac{d^2\vek{z}_1}{(2\pi)^2}\, \frac{d^2\vek{z}_2}{(2\pi)^2}\;
    e^{-i \vek{z}_1 \vek{q}{}_1 -i \vek{z}_2 \vek{q}{}_2}
  \int d^2\vek{y}\; \vek{y}^2 M^2
\nonumber \\
& \quad \times \Bigl(
\bigl[ (1+\cos^2\theta_1) K_{q_1 \bar{q}_3}
       + 2 \cos\theta_1\, \Kp{q_1 \bar{q}_3} \bigr] 
\nonumber \\
& \qquad \times \Bigl\{ \bigl[\,
   \cos 2(\varphi_2 - \varphi_y)\,  K_{\delta q_2 \delta \bar{q}_2}
 - \sin 2(\varphi_2 - \varphi_y)\, \Kp{\delta q_2 \delta \bar{q}_2} \bigr]
\nonumber \\
\phantom{\Bigl[ \Bigr]}
& \qquad\qquad \times
 ( f_{q_1 \delta q_2} \bar{f}_{\bar{q}_3 \delta\bar{q}_2}
 - g_{q_1 \delta q_2} \bar{g}_{\bar{q}_3 \delta\bar{q}_2}
 - f_{\Delta q_1 \delta q_2} \bar{f}_{\Delta\bar{q}_3 \delta\bar{q}_2}
 + g_{\Delta q_1 \delta q_2} \bar{g}_{\Delta\bar{q}_3 \delta\bar{q}_2} )
\nonumber \\
\phantom{\Bigl[ \Bigr]}
& \qquad \hspace{0.6em} + \bigl[
   \sin 2(\varphi_2 - \varphi_y)\,  K_{\delta q_2 \delta \bar{q}_2}
 + \cos 2(\varphi_2 - \varphi_y)\, \Kp{\delta q_2 \delta \bar{q}_2} \bigr]
\nonumber \\
\phantom{\Bigl[ \Bigr]}
& \qquad\qquad \times
 ( f_{q_1\delta q_2} \bar{g}_{\bar{q}_3\delta\bar{q}_2}
 + g_{q_1\delta q_2} \bar{f}_{\bar{q}_3\delta\bar{q}_2}
 + f_{\Delta q_1\delta q_2} \bar{g}_{\Delta\bar{q}_3\delta\bar{q}_2}
 + g_{\Delta q_1\delta q_2} \bar{f}_{\Delta\bar{q}_3\delta\bar{q}_2} ) \Bigr\}
\nonumber \\
\phantom{\Bigl[ \Bigr]}
& \quad \hspace{0.6em} -
\bigl[ (1+\cos^2\theta_1) K_{q_1 \Delta \bar{q}_3}
      + 2 \cos\theta_1\, \Kp{q_1\Delta \bar{q}_3} \bigr]
\nonumber \\
& \qquad \times \Bigl\{ \bigl[\,
   \cos 2(\varphi_2 - \varphi_y)\,  K_{\delta q_2 \delta \bar{q}_2}
 - \sin 2(\varphi_2 - \varphi_y)\, \Kp{\delta q_2 \delta \bar{q}_2} \bigr]
\nonumber \\
\phantom{\Bigl[ \Bigr]}
& \qquad\qquad \times
 ( f_{q_1 \delta q_2} \bar{g}_{\Delta\bar{q}_3 \delta\bar{q}_2}
 - g_{q_1 \delta q_2} \bar{f}_{\Delta\bar{q}_3 \delta\bar{q}_2}
 - f_{\Delta q_1 \delta q_2} \bar{g}_{\bar{q}_3 \delta\bar{q}_2}
 + g_{\Delta q_1 \delta q_2} \bar{f}_{\bar{q}_3 \delta\bar{q}_2} )
\nonumber \\
\phantom{\Bigl[ \Bigr]}
& \qquad \hspace{0.6em} + \bigl[\,
   \sin 2(\varphi_2 - \varphi_y)\, K_{\delta q_2 \delta \bar{q}_2}
 + \cos 2(\varphi_2 - \varphi_y)\, \Kp{\delta q_2 \delta \bar{q}_2} \bigr]
\nonumber \\
& \qquad\qquad \times
 ( f_{q_1\delta q_2} \bar{f}_{\Delta\bar{q}_3\delta\bar{q}_2}
 + g_{q_1\delta q_2} \bar{g}_{\Delta\bar{q}_3\delta\bar{q}_2}
 + f_{\Delta q_1\delta q_2}\bar{f}_{\bar{q}_3\delta\bar{q}_2}
 + g_{\Delta q_1\delta q_2}\bar{g}_{\bar{q}_3\delta\bar{q}_2} ) \Bigr\}
\Bigr)
\nonumber \\
\phantom{\Bigl[ \Bigr]}
& \quad + \{ \text{flavor interference} \}
        + \{ q \bar{q}~\text{permutations} \}
        + \{ \text{transv.\ pol.\ in interaction 1} \} \,,
\end{align}
where the flavor interference and $q\bar{q}$ permutation terms are
obtained in the same way as in \eqref{eq:zeroT}.  The terms for transverse
polarization in interaction 1 are obtained by replacing labels as $1 \to
2$, $2\to 1$, $3 \to 4$ in the coupling factors and by making the same
replacement in the DPD subscripts after interchanging their order, i.e.\
$\smash{f_{a_1 \delta q_2} \rightarrow f_{\delta q_1 a_2}}$,
$\smash{\bar{g}_{\bar{a}_3 \delta\bar{q}_2} \rightarrow
  \bar{g}_{\delta\bar{q}_1 \bar{a}_4}}$~etc.

The azimuthal angle $\varphi_2$ of the lepton produced in interaction 2
has already been defined, and $\varphi_y$ is the azimuthal angle of
$\vek{y}$ in the $pp$ center-of-mass.  As anticipated in
section~\ref{sec:boson-decay-frames}, the cross section depends only on
the difference $\varphi_2 - \varphi_y$ of these angles, in agreement with
rotation invariance.  The $\varphi_y$ dependence in \eqref{eq:oneT} arises
from the uncontracted vectors $\vek{y}$ and $\tilde{\vek{y}}$ in the DPDs
\eqref{eq:DPD2} for transversely polarized partons: it is hence this
polarization which enables a dependence of the cross section on the
azimuthal angle of the produced lepton.

The transverse distance $\vek{y}$ is integrated over in \eqref{eq:oneT}
and hence not measurable.  The $\vek{y}$ integration is nontrivial because
the DPDs depend on the azimuthal angles between $\vek{y}$ and $\vek{z}_1$
and $\vek{z}_2$, whose directions are in turn correlated with those of
$\vek{q}_1$ and $\vek{q}_2$ through the exponential $e^{-i \vek{z}_1
  \vek{q}{}_1 -i \vek{z}_2 \vek{q}{}_2}$.  The integral over $\vek{y}$,
$\vek{z}_1$ and $\vek{z}_2$ in the cross section thus turns the
$\varphi_y$ dependence into a dependence on the azimuthal angles of the
transverse momenta $\vek{q}_1$ and $\vek{q}_2$.  All together we thus see
that a correlation between $\vek{y}$ and the transverse polarization of
parton 2 in the DPDs leads to an azimuthal correlation between the lepton
from interaction 2 and both transverse vector boson momenta.  This is
similar (but not identical) to single Drell-Yan production, where a
correlation between the transverse polarization of a parton and its
transverse momentum induces an azimuthal correlation between the momenta
of the vector boson and its decay lepton \cite{Boer:1999mm}.

We finally turn to the case where both vector bosons are produced from
transversely polarized quarks.  The corresponding contribution to the
cross section is
\begin{align}
  \label{eq:doubleT}
& \frac{d\sigma^{(2)}}{\prod_{i=1}^2
   dx_i\, d\bar{x}_i\, d^2\vek{q}{}_i\, d\Omega_i}
= \frac{1}{C}\; 2 \sin^2\theta_1\, \sin^2\theta_2\,
  \sum_{q_1q_2}
  \int\frac{d^2\vek{z}_1}{(2\pi)^2}\, \frac{d^2\vek{z}_2}{(2\pi)^2}\;
    e^{-i \vek{z}_1 \vek{q}{}_1 -i \vek{z}_2 \vek{q}{}_2}
  \int d^2\vek{y}
\nonumber \\
& \quad \times \Bigl\{ \bigl[\, \cos 2(\varphi_1 - \varphi_2)\,
 ( K_{\delta q_1 \delta \bar{q}_1}  K_{\delta q_2 \delta \bar{q}_2}
+ \Kp{\delta q_1 \delta \bar{q}_1} \Kp{\delta q_2 \delta \bar{q}_2} )
 \nonumber \\
& \qquad\quad - \sin 2(\varphi_1 - \varphi_2)\,
 ( \Kp{\delta q_1 \delta \bar{q}_1}  K_{\delta q_2 \delta \bar{q}_2}
  - K_{\delta q_1 \delta \bar{q}_1} \Kp{\delta q_2 \delta \bar{q}_2} )
\bigr]
\nonumber \\
& \qquad\qquad \times
( f_{\delta q_1 \delta q_2} \bar{f}_{\delta\bar{q}_1 \delta\bar{q}_2}
 - \vek{y}^4 M^4 g_{\delta q_1 \delta q_2}^{a}
           \bar{g}_{\delta\bar{q}_1 \delta\bar{q}_2}^{a} )
\phantom{\Bigl[ \Bigr]}
\nonumber \\
& \qquad + \bigl[\, \sin 2(\varphi_1 - \varphi_2)\,
 ( K_{\delta q_1 \delta \bar{q}_1}  K_{\delta q_2 \delta \bar{q}_2}
+ \Kp{\delta q_1 \delta \bar{q}_1} \Kp{\delta q_2 \delta \bar{q}_2} )
\phantom{\Bigl[ \Bigr]}
\nonumber \\
& \qquad\quad + \cos 2(\varphi_1 - \varphi_2)\,
( \Kp{\delta q_1 \delta \bar{q}_1}  K_{\delta q_2 \delta \bar{q}_2}
 - K_{\delta q_1 \delta \bar{q}_1} \Kp{\delta q_2 \delta \bar{q}_2} )
\bigr] 
\nonumber \\
& \qquad\qquad \times \vek{y}^2 M^2\,
( f_{\delta q_1 \delta q_2} \bar{g}_{\delta\bar{q}_1 \delta\bar{q}_2}^{a}
+ g_{\delta q_1 \delta q_2}^{a} \bar{f}_{\delta\bar{q}_1 \delta\bar{q}_2} )
\phantom{\Bigl[ \Bigr]}
\nonumber \\
& \qquad + \bigl[\, \cos 2(\varphi_1 + \varphi_2 - 2\varphi_y)\,
 ( K_{\delta q_1 \delta \bar{q}_1}  K_{\delta q_2 \delta \bar{q}_2}
- \Kp{\delta q_1 \delta \bar{q}_1} \Kp{\delta q_2 \delta \bar{q}_2} )
\phantom{\Bigl[ \Bigr]}
\nonumber \\
& \qquad\quad - \sin 2(\varphi_1 + \varphi_2 - 2\varphi_y)\,
( \Kp{\delta q_1 \delta \bar{q}_1}  K_{\delta q_2 \delta \bar{q}_2}
 + K_{\delta q_1 \delta \bar{q}_1} \Kp{\delta q_2 \delta \bar{q}_2} )
\bigr] 
\nonumber \\
& \qquad\qquad \times \vek{y}^4 M^4\,
( f_{\delta q_1\delta q_2}^{t} \bar{f}_{\delta\bar{q}_1\delta\bar{q}_2}^{t}
- g_{\delta q_1\delta q_2}^{s}\bar{g}_{\delta\bar{q}_1\delta\bar{q}_2}^{s} )
\phantom{\Bigl[ \Bigr]}
\nonumber \\
& \qquad - \bigl[\, \sin 2(\varphi_1 + \varphi_2 - 2\varphi_y)\,
 ( K_{\delta q_1 \delta \bar{q}_1}  K_{\delta q_2 \delta \bar{q}_2}
- \Kp{\delta q_1 \delta \bar{q}_1} \Kp{\delta q_2 \delta \bar{q}_2} )
\phantom{\Bigl[ \Bigr]}
\nonumber \\
& \qquad\quad + \cos 2(\varphi_1 + \varphi_2 - 2\varphi_y)\,
( \Kp{\delta q_1 \delta \bar{q}_1}  K_{\delta q_2 \delta \bar{q}_2}
 + K_{\delta q_1 \delta \bar{q}_1} \Kp{\delta q_2 \delta \bar{q}_2} )
\bigr]
\nonumber \\
& \qquad\qquad \times \vek{y}^4 M^4\,
( f_{\delta q_1\delta q_2}^{t} \bar{g}_{\delta\bar{q}_1\delta\bar{q}_2}^{s}
+ g_{\delta q_1\delta q_2}^{s} \bar{f}_{\delta\bar{q}_1\delta\bar{q}_2}^{t} )
\Bigr\}
\nonumber \\
& \quad + \{ \text{flavor interference} \}
        + \{ q \bar{q}~\text{permutations} \}
\phantom{\Bigl[ \Bigr]}
\end{align}
and depends on the azimuthal angles $\varphi_1$, $\varphi_2$ and
$\varphi_y$ in addition to the polar angles $\theta_1$ and $\theta_2$.
The flavor interference and $q\bar{q}$ permutation terms are again
obtained as in \eqref{eq:zeroT}.

The terms depending on $\varphi_1 - \varphi_2$ describe a transverse
correlation between the leptonic decay planes of the vector bosons.  By
contrast, the terms with $\varphi_1 + \varphi_2 - 2\varphi_y$ describe an
azimuthal correlation between the lepton momenta and the direction between
the hard interactions, which after integration over $\vek{y}$, $\vek{z}_1$
and $\vek{z}_2$ turns into an azimuthal correlation between the lepton
momenta and the momenta of the two bosons.


\subsection{Cross section integrated over transverse boson momenta}

Integration over the transverse momenta of the two vector bosons yields
cross sections expressed in terms of collinear double parton distributions
$F(x_1,x_2,\vek{y})$.  Their spin structure is as in \eqref{eq:DPD1} to
\eqref{eq:DPD3} but without pseudoscalar functions $g$, because one cannot
construct a pseudoscalar with only one vector $\vek{y}$.

Upon integration over $\vek{q}_1$ and $\vek{q}_2$, the cross section
\eqref{eq:cross-WW} for $W$ pair production becomes
\begin{align}
  \label{eq:coll-WW}
& \frac{d\sigma^{WW}}{\prod_{i=1}^2
   dx_i\, d\bar{x}_i\, d\Omega_i}
= \frac{1}{C}\,
  \sum_{q_1 q_2 q_3 q_4}
  K_{q_1\bar{q}_3} K_{q_2\bar{q}_4}
\nonumber \\
& \quad \times \biggl\{
 (1 + \cos\theta_1)^2 \, (1 + \cos\theta_2)^2
\int d^2\vek{y}\;
 (f_{q_1 q_2} + f_{\Delta q_1 \Delta q_2})
 (\bar{f}_{\bar{q}_3\bar{q}_4} + \bar{f}_{\Delta\bar{q}_3\Delta \bar{q}_4})
\nonumber \\
& \qquad \!\bs +
 (1 + \cos\theta_1)^2 \, (1 - \cos\theta_2)^2
\int d^2\vek{y}\;
 (f_{q_1 \bar{q}_4} - f_{\Delta q_1 \Delta \bar{q}_4})
 (\bar{f}_{\bar{q}_3 {q}_2} - \bar{f}_{\Delta\bar{q}_3 \Delta {q}_2})
\nonumber \\
& \qquad \!\bs +
 (1 - \cos\theta_1)^2 \, (1 + \cos\theta_2)^2
\int d^2\vek{y}\;
 (f_{\bar{q}_3 q_2} - f_{\Delta \bar{q}_3 \Delta q_2})
 (\bar{f}_{{q}_1 \bar{q}_4} - \bar{f}_{\Delta {q}_1 \Delta \bar{q}_4})
\nonumber \\
& \qquad \!\bs +
 (1 - \cos\theta_1)^2 \, (1 - \cos\theta_2)^2
\int d^2\vek{y}\;
 (f_{\bar{q}_3 \bar{q}_4} + f_{\Delta \bar{q}_3 \Delta \bar{q}_4})
 (\bar{f}_{{q}_1 {q}_2} + \bar{f}_{\Delta {q}_1 \Delta {q}_2})
\biggr\}
\nonumber \\
& \quad + \{ \text{flavor interference} \} \,,
  \phantom{\int}
\end{align}
where the arguments of the distributions are $f(x_1,x_2,\vek{y})$ and
$\bar{f}(\bar{x}_1,\bar{x}_2,\vek{y})$.
In the general case we have a contribution
\begin{align}
  \label{eq:coll-zero}
&\frac{d\sigma^{(0)}}{\prod_{i=1}^2
   dx_i\, d\bar{x}_i\, d\Omega_i} =
   \frac{1}{C}\, \sum_{q_1 q_2 q_3 q_4}
\nonumber \\
& \quad \times \biggl\{ 
  \left[ (1+\cos^2\theta_1) K_{q_1 \bar{q}_3}
        + 2 \cos\theta_1\, \Kp{q_1 \bar{q}_3} \right]
  \left[ (1+\cos^2\theta_2) K_{q_2 \bar{q}_4}
        + 2 \cos\theta_2\, \Kp{q_2 \bar{q}_4} \right] 
\nonumber \\[-0.3em]
& \qquad\quad \times \int d^2\vek{y}\;
  (f_{q_1 q_2} \bar{f}_{\bar{q}_3 \bar{q}_4}
 + f_{\Delta q_1 \Delta q_2} \bar{f}_{\Delta\bar{q}_3 \Delta\bar{q}_4})
\nonumber \\[0.2em]
& \qquad
 + \left[ (1+\cos^2\theta_1) K_{q_1 \Delta\bar{q}_3}
         + 2 \cos\theta_1\, \Kp{q_1 \Delta\bar{q}_3} \right]
   \left[ (1+\cos^2\theta_2) K_{q_2 \Delta\bar{q}_4}
         + 2 \cos\theta_2\, \Kp{q_2 \Delta\bar{q}_4} \right]
\nonumber \\
& \qquad\quad \times \int d^2\vek{y}\;
  (f_{q_1 q_2} \bar{f}_{\Delta\bar{q}_3 \Delta\bar{q}_4}
 + f_{\Delta q_1 \Delta q_2} \bar{f}_{\bar{q}_3 \bar{q}_4})
\biggr\}
\nonumber \\
& \quad + \left\{ \text{flavor interference} \right\}
        + \{ q \bar{q}~\text{permutations} \}
\phantom{\Bigl[ \Bigr]}
\end{align}
from unpolarized and longitudinally polarized partons.  The contribution
with transverse quark polarization in one of the two hard interactions now
vanishes,
\begin{align}
\frac{d\sigma^{(1)}}{\prod_{i=1}^2
   dx_i\, d\bar{x}_i\, d\Omega_i} = 0 \,.
\end{align}
This is because integration of \eqref{eq:oneT} over $\vek{q}_1$ and
$\vek{q}_2$ sets $\vek{z}_1 = \vek{z}_2 = \vek{0}$, after which the
$\vek{y}$ integral gives zero due to the azimuthal dependence on
$\varphi_y$.  By contrast, the contribution with transverse quark
polarization in both hard interactions remains nonzero,
\begin{align}
& \frac{d\sigma^{(2)}}{\prod_{i=1}^2
   dx_i\, d\bar{x}_i\, d\Omega_i} = \frac{1}{C}\;
  2 \sin^2\theta_1\, \sin^2\theta_2\, \sum_{q_1 q_2}
\nonumber \\
& \quad \times \biggl\{ \bigl[\, \cos2(\varphi_1-\varphi_2)\,
  ( K_{\delta q_1 \delta \bar{q}_1}  K_{\delta q_2 \delta \bar{q}_2}
 + \Kp{\delta q_1 \delta \bar{q}_1} \Kp{\delta q_2 \delta \bar{q}_2} ) 
\nonumber \\[-0.4em]
& \qquad\quad - \sin2(\varphi_1-\varphi_2)\,
  ( \Kp{\delta q_1 \delta \bar{q}_1}  K_{\delta q_2 \delta \bar{q}_2}
   - K_{\delta q_1 \delta \bar{q}_1} \Kp{\delta q_2 \delta \bar{q}_2} ) \bigr]
\int d^2\vek{y}\;
    f_{\delta q_1 \delta q_2} \bar{f}_{\delta\bar{q}_1 \delta\bar{q}_2}
\biggr\}
\nonumber \\
& \quad + \{ \text{flavor interference} \}
        + \{ q \bar{q}~\text{permutations} \} \,.
\end{align}
According to \eqref{eq:DPD3} the distribution $f_{\delta q_1\delta q_2}$
describes the correlation between the directions of the transverse
polarizations of two quarks in the proton.  This correlation and its
counterpart for antiquarks induce a correlation between the leptonic decay
planes of the vector bosons, even if their transverse momenta are
integrated over.  Only if one integrates over the azimuthal angle of at
least one of the leptons does the contribution from transverse quark
polarization completely disappear from the cross section.

The cross section of the double Drell-Yan process with two photons was
calculated in \cite{Manohar:2012jr}, integrated over the transverse boson
momenta and over the angles of the decay leptons.  The expression in
equation~(9) of \cite{Manohar:2012jr} agrees with our result
\eqref{eq:coll-zero} (up to the combinatorial factor $1/C$, which was
omitted in \cite{Manohar:2012jr}).


\section{Transverse position dependence of distributions}
\label{sec:model}

So far we have focused on spin correlations in DPDs and their consequences
for the double Drell-Yan cross section.  Even for unpolarized partons,
however, there can be correlations between two partons in the proton,
namely correlations affecting the dependence of DPDs on the transverse
variables $\vek{y}$, $\vek{z}_1$ and $\vek{z}_2$ and the interplay between
these variables and the longitudinal momentum fractions.

In the present section we take a brief look at this issue by using a
simple model in which the proton is described by a three-quark wave
function.  This is clearly too simple to describe the physics of small
momentum fractions most relevant at the LHC, although it may actually be
used for modeling quark DPDs at momentum fractions in the valence
region. We proceed with this model in the spirit of an exploratory study.

Our model ansatz for the three-quark light-cone wave function of the
proton is
\begin{align}
  \label{eq:lc-wf}
\Psi(x_i, \vek{b}_i - \vek{b}) = \Phi(x_i)\, 
  \exp\biggl[- \frac{1}{4a^2} \sum_{i=1}^3
      x_i (\vek{b}_i - \vek{b})^2 \biggr] \,,
\end{align}
where $a$ is parameter of dimension length, $\vek{b} = x_1 \vek{b}_1 + x_2
\vek{b}_2 + x_3 \vek{b}_3$ is the transverse position of the proton, and
$x_1 + x_2 + x_3 = 1$.  The corresponding wave function depending on
transverse momenta is a Gaussian with exponent $- a^2 \sum_i \vek{k}_i^2
/x_i^{}$, which was long ago proposed in \cite{Brodsky:1980vj} and is
often used for the phenomenology of valence dominated quantities, see
e.g.~\cite{Diehl:1998kh}.  The relation between the light-cone wave
functions in transverse momentum and transverse position representation
can be found in \cite{Diehl:2003ny}.  We do not specify the longitudinal
part $\Phi(x_i)$ of the wave function nor its spin-flavor dependence here,
since the focus of our study is on the transverse variables.

From the light-cone wave function \eqref{eq:lc-wf} one can compute the
contribution of the three-quark Fock state to the DPD of two quarks in the
proton, in full analogy to the well-known case of single-parton
distributions (discussed e.g.\ in \cite{Diehl:2003ny}).  Up to a factor
depending only on the longitudinal momentum fractions $x_i$, the DPD is
given by
\begin{align}
  \label{eq:transDPD}
F(x_i,\vek{z}_i,\vek{y}) & \propto
\exp\biggl[- \frac{1}{8a^2}\,
  \biggl\{ x_1 (1-x_1) \vek{z}_1^2 - 2 x_1 x_2 \vek{z}_1 \vek{z}_2
       + x_2 (1-x_2) \vek{z}_2^2
       + \frac{4x_1 x_2}{x_1 + x_2}\ms \vek{y}^2 \biggr\} \biggr]
\nonumber \\
 & \quad\times \int d^2\vek{b}\, \exp\biggl[ - \frac{1}{2a^2}\,
   \frac{x_1 + x_2}{1 -  x_1 - x_2}\,
   \Bigl( \vek{b} + \frac{x_1}{x_1 + x_2} \vek{y} \Bigr)^2 \biggr] \,,
\end{align}
where $\vek{b}$ is the transverse position of the proton, averaged over
the scattering amplitude and its conjugate as specified in
\cite{Diehl:2011yj}.  The second line in \eqref{eq:transDPD} just gives an
$x_i$ dependent factor after integration over $\vek{b}$.

Inserting \eqref{eq:transDPD} into the cross section formula
\eqref{eq:cross-simple} and performing the integrals over all transverse
positions, one obtains a cross section for double hard scattering that
depends on the transverse boson momenta as
\begin{align}
  \label{eq:trans}
\exp\biggl\{ - a^2 \Bigl[\, \vek{q}_1^2\, C_{11}^{}(x_i,\bar{x}_i)
  + 2 \vek{q}_1 \vek{q}_2\, C_{12}^{}(x_i,\bar{x}_i)
  + \vek{q}_2^2\, C_{22}^{}(x_i,\bar{x}_i) \,\Bigr] \biggr\}
\end{align}
with dimensionless functions $C_{ij}$ of the momentum fractions $x_1,x_2$
and $\bar{x}_1,\bar{x}_2$, which are somewhat lengthy and will not be
given here.  The expression in square brackets is positive
definite, so that the transverse momentum dependence has a Gaussian
falloff at large transverse momenta.  The coefficient $C_{12}$ describing
the correlation between $\vek{q}_1$ and $\vek{q}_2$ is positive as well,
so that one finds a preference for the two vector bosons to have opposite
transverse momenta.  We see that even with the simple wave function ansatz
\eqref{eq:lc-wf} the dependence of the cross section on the transverse
momenta of the gauge bosons is not independent of their longitudinal
momenta.

An ansatz often made in phenomenology is to neglect correlations between
partons and to write a collinear DPD as the convolution of two
single-parton distributions that depend on the momentum fraction and the
transverse position of the parton.  This ansatz can be extended to include
$\vek{z}_i$ dependent DPDs and then reads
\begin{align}
  \label{eq:no-corr}
F(x_i,\vek{z}_i,\vek{y}) & \approx  \int d^2\vek{b}\;
  f\bigl(x_2,\vek{z}_2; \vek{b} + \half x_1 \vek{z}_1 \bigr)\,
  f\bigl(x_1,\vek{z}_1; \vek{b} + \vek{y} - \half x_2 \vek{z}_2 \bigr) \,.
\end{align}
The second argument of the single-parton distribution $f$ is Fourier
conjugate to the transverse quark momentum and the third argument gives
the transverse position of the proton with respect to the quark, both
averaged over the scattering amplitude and its conjugate.  The shift of
this argument by $\half x_1 \vek{z}_1$ or $-\half x_2 \vek{z}_2$ is a
consequence of Lorentz invariance as explained in \cite{Diehl:2011tt}.
Evaluating the single-parton distributions for the light-cone wave
function \eqref{eq:lc-wf} one obtains
\begin{align}
  \label{eq:transDPD-no-corr}
F(x_i,\vek{z}_i,\vek{y}) & \propto \int d^2\vek{b}\,
  \exp\biggl[- \frac{1}{8a^2}\,
    \frac{x_2}{1-x_2}\, \Bigl\{ (1-x_2)^2 \vek{z}_2^2
           + (2\vek{b} + x_1 \vek{z}_1)^2 \Bigr\} \biggr]
\nonumber \\
& \qquad\quad \times 
  \exp\biggl[- \frac{1}{8a^2}\,
    \frac{x_1}{1-x_1}\, \Bigl\{ (1-x_1)^2 \vek{z}_1^2
           + (2\vek{b} + 2\vek{y} - x_2 \vek{z}_2)^2 \Bigr\} \biggr]
\nonumber \\
&\propto \exp\biggl[- \frac{1}{8a^2}\, \biggl\{
   x_1 (1-x_1) \vek{z}_1^2 + x_2 (1-x_2) \vek{z}_2^2
\nonumber \\[-0.4em]
&\qquad\qquad\qquad + \frac{x_1 x_2}{x_1 (1-x_2) + (1-x_1) x_2}
     \Bigl( 2 \vek{y} - x_1 \vek{z}_1 - x_2 \vek{z}_2 \Bigr)^2
 \biggr\} \biggr]
\end{align}
for the transverse dependence of the DPD.  This is visibly different from
the result \eqref{eq:transDPD} of the direct calculation.  Although the
ansatz \eqref{eq:no-corr} involves the convolution of two single-parton
distributions, thus suggesting that the two partons are distributed
independently, it induces correlations between transverse and longitudinal
variables in $F(x_i,\vek{z}_i,\vek{y})$.

Inserting the form \eqref{eq:transDPD-no-corr} into the cross section
formula one obtains again a Gaussian behavior as in \eqref{eq:trans}, but
with different coefficients $C_{ij}$.  In particular, the sign of $C_{12}$
is then equal to the sign of $(x_1-\bar{x}_1)(x_2-\bar{x}_2)$, so that
depending on the longitudinal momentum fractions the transverse boson
momenta tend to be in the same hemisphere or in opposite ones.  This
difference in qualitative behavior shows that the ansatz
\eqref{eq:no-corr} must be used with great care when one is interested in
correlation effects.

Setting $\vek{z}_1 = \vek{z}_2 = \vek{0}$ in \eqref{eq:transDPD} and
\eqref{eq:transDPD-no-corr} gives collinear DPDs with a Gaussian
dependence on $\vek{y}$.  The Gaussian width depends on $x_1$ and $x_2$
and differs in the two cases,
\begin{align}
F(x_i, \vek{y}) \Big|_{\protect\eqref{eq:transDPD}}
& \propto \exp\biggl[- \frac{1}{2a^2}\,
     \frac{x_1 x_2}{x_1 + x_2}\, \vek{y}^2 \biggr] \,,
\nonumber \\
F(x_i, \vek{y})\Big|_{\protect\eqref{eq:transDPD-no-corr}}
& \propto \exp\biggl[- \frac{1}{2a^2}\,
     \frac{x_1 x_2}{x_1 + x_2 - 2 x_1 x_2}\, \vek{y}^2 \biggr] \,.
\end{align}
We see that, within our model, the ansatz \eqref{eq:no-corr} does not
reproduce the interplay between $\vek{y}$ and the momentum fractions.  It
does, however, provide a valid approximation unless $x_1$ and $x_2$ are
both rather large.


\section{Summary}
\label{sec:summary}

Multiple hard interactions in $pp$ collisions can yield substantial
contributions to the production of final states with high multiplicity in
parts of phase space.  In this paper we have shown how spin correlations
between two partons in the proton affect the rate and the angular
distribution of the final state in the production of four leptons via two
electroweak gauge bosons.  We considered both the case where the
transverse momenta of the bosons are small (using transverse-momentum
dependent factorization) and the case where they are integrated over
(using collinear factorization).

We find that longitudinal spin correlations between the quarks or
antiquarks in the proton affect the rate of double parton scattering, and
in the presence of axial-vector currents also the polar distribution
of the produced leptons.  Correlations involving transversely polarized
quarks or antiquarks induce azimuthal correlations between the final state
leptons.  A part of these correlations persists if the transverse momenta
of the gauge bosons are integrated over.  Having two ``independent'' hard
interactions in double parton scattering does hence not imply that the
final states produced by the two interactions are independent of each
other.  

How large parton spin correlations inside a proton actually are remains an
open question that deserves further study.  This also holds for possible
correlations between the transverse distribution and the longitudinal
momentum fractions of the partons.  We find several such correlations in a
simple model with a three-quark wave function.  Within this model we also
find that the often used ansatz to represent double parton distributions
as convolutions of single-parton distributions is inadequate to describe
details of the kinematic dependence in double parton scattering.

Double Drell-Yan production involves a particularly simple hard-scattering
subprocess but nevertheless exhibits a rich pattern of angular effects
induced by parton spin correlations.  It is natural to expect that other
processes, in particular those involving multijets, will share this
feature.  We note that the cross section dependence on angles between the
final-state particles implies a dependence on the invariant mass of
particle pairs, which is an important quantity in searches for new
physics.  An estimate of the possible size of such effects would therefore
be of great value.


\section*{Acknowledgments}

It is our pleasure to thank J.~Gaunt and L.~Zeune for valuable remarks on
the manuscript.


\appendix
\section{Coupling Factors}
\label{sec:coupling-factors}

In this appendix we list the coupling factors $K$ and $K'$ appearing in
the double Drell-Yan cross section.  Further relations between these
factors are given in section~\ref{sec:hard-proc}.

\subsection{Charged vector bosons}

For $W^+$ production one has
\begin{align}
K_{q_i \bar{q}_j} & =
  \frac{\alpha^2}{4 N_c}\, \frac{|V_{q_i q_j}|^2}{(2\ms \sin \theta_w)^4}\,
  \frac{Q_i^2}{(Q_i^2 - m_W^2)^2 + m_W^2 \Gamma_W^2} \,,
\qquad (e_{q_i} - e_{q_j} = 1) \,,
\intertext{and for $W^-$ production}
K_{q_i \bar{q}_j} & =
  \frac{\alpha^2}{4 N_c}\, \frac{|V_{q_j q_i}|^2}{(2\ms \sin \theta_w)^4}\,
  \frac{Q_i^2}{(Q_i^2 - m_W^2)^2 + m_W^2 \Gamma_W^2} \,,
\qquad (e_{q_i} - e_{q_j} = -1) \,.
\end{align}
Here $N_c = 3$ is the number of colors, $V_{q_i q_j}$ a CKM matrix
element, $\theta_w$ the weak mixing angle, $\alpha$ the electromagnetic
fine structure constant, and $e_{q_i}$ the charge of quark $q_i$ in units
of the positron charge.


\subsection{Neutral vector bosons}

For a lepton pair $\ell^+ \ell^-$ produced via a $\gamma^*$, $Z$ or their
interference, one has coupling factors
\begin{align}
K_{q_i \bar{q}_j} = \frac{\alpha^2}{4 N_c} & \biggl\{
  \frac{e_{q_i} e_{q_j}}{Q_i^2}
  - A(Q_i)\, g^V_{\ell} (e_{q_i} g^V_{q_j} + e_{q_j} g^V_{q_i})
- i B(Q_i)\, g^V_{\ell} (e_{q_i} g^V_{q_j} - e_{q_j} g^V_{q_i})
\nonumber \\
& + C(Q_i)\, \big[ (g^V_{\ell})^2 + (g^A_{\ell})^2 \big]
                   (g^V_{q_i} g^V_{q_j}+ g^A_{q_j} g^A_{q_i}) \biggr\} \,,
\nonumber \\
\Kp{q_i \bar{q}_j} =  \frac{\alpha^2}{4 N_c} & \biggl\{
  - A(Q_i)\, g^A_{\ell} (e_{q_i} g^A_{q_j} + e_{q_j} g^A_{q_i})
- i B(Q_i)\, g^A_{\ell} (e_{q_i} g^A_{q_j} - e_{q_j} g^A_{q_i})
\nonumber \\
& + C(Q_i)\, 2 g^V_{\ell} g^A_{\ell}
             (g^V_{q_i} g^A_{q_j} + g^V_{q_j} g^A_{q_i}) \biggr\} \,,
\nonumber \\
K_{q_i \Delta\bar{q}_j} =  \frac{\alpha^2}{4 N_c} & \biggl\{
  - A(Q_i)\, g^V_{\ell} (e_{q_i} g^A_{q_j} + e_{q_j} g^A_{q_i})
- i B(Q_i)\, g^V_{\ell} (e_{q_i} g^A_{q_j} - e_{q_j} g^A_{q_i})
\nonumber \\
& + C(Q_i)\, \big[ (g^V_{\ell})^2 + (g^A_{\ell})^2 \big]
                   (g^V_{q_i} g^A_{q_j} + g^V_{q_j} g^A_{q_i}) \biggr\} \,,
\nonumber \\
\Kp{q_i \Delta\bar{q}_j} =  \frac{\alpha^2}{4 N_c} & \biggl\{
 -  A(Q_i)\, g^A_{\ell} (e_{q_i} g^V_{q_j} + e_{q_j} g^V_{q_i})
 - i B(Q_i)\, g^A_{\ell} (e_{q_i} g^V_{q_j} - e_{q_j} g^V_{q_i})
\nonumber \\
& + C(Q_i)\, 2 g^V_{\ell} g^A_{\ell}
             (g^V_{q_i} g^V_{q_j} + g^A_{q_j} g^A_{q_i}) \biggr\}
\intertext{and}
K_{\delta q_i \delta\bar{q}_j} =  \frac{\alpha^2}{4 N_c} & \biggl\{
  \frac{e_{q_i} e_{q_j}}{Q_i^2}
  - A(Q_i)\, g^V_{\ell} (e_{q_i} g^V_{q_j} + e_{q_j} g^V_{q_i})
- i B(Q_i)\, g^V_{\ell} (e_{q_i} g^V_{q_j} - e_{q_j} g^V_{q_i})
\nonumber \\
& + C(Q_i)\, \big[ (g^V_{\ell})^2 + (g^A_{\ell})^2 \big]
                   (g^V_{q_i} g^V_{q_j} - g^A_{q_j} g^A_{q_i}) \biggr\} \,,
\nonumber \\
\Kp{\delta q_i\delta \bar{q}_j} =  \frac{\alpha^2}{4 N_c} & \biggl\{
 -  B(Q_i)\, g^V_{\ell} (e_{q_i} g^A_{q_j} + e_{q_j} g^A_{q_i})
+ i A(Q_i)\, g^V_{\ell} (e_{q_i} g^A_{q_j} - e_{q_j} g^A_{q_i})
\nonumber \\
& - i C(Q_i)\, \big[ (g^V_{\ell})^2 + (g^A_{\ell})^2 \big]
             (g^V_{q_i} g^A_{q_j} - g^V_{q_j} g^A_{q_i}) \biggr\} \,.
\end{align}
Here we have used the conventional vector and axial fermion couplings to
the $Z$ boson,
\begin{align}
g^V_f &= I_{f}^3 - 2 e_f \sin^2\theta_w \,,
&
g^A_f &= I_{f}^3 \,,
\end{align}
where $I_{f}^3$ is the third component of the weak isospin of the left
handed fermion $f$ and $e_f$ its charge in units of positron charge.
Since we do not consider $Z$ decays to neutrinos, $\ell$ is always a
negatively charged lepton.  We have furthermore used the abbreviations
\begin{align}
A(Q_i) &= \frac{1}{\sin^2 2 \theta_w}\,
          \frac{Q_i^2 - m_Z^2}{(Q_i^2 - m_Z^2)^2 + m_Z^2 \Gamma_Z^2} \,,
&
B(Q_i) &= \frac{1}{\sin^2 2 \theta_w}\,
          \frac{m_Z \Gamma_Z}{(Q_i^2 - m_Z^2)^2 + m_Z^2 \Gamma_Z^2} \,,
\nonumber \\
C(Q_i) &= \frac{1}{\sin^4 2 \theta_w}\,
          \frac{Q_i^2}{(Q_i^2 - m_Z^2)^2 + m_Z^2 \Gamma_Z^2} \,.
\end{align}
For the usual hard-scattering cross sections one has equal flavors $q_i =
q_j$ in the above coupling factors and finds that their imaginary parts
are zero.  This is not the case for the coupling factors describing flavor
interference, where $q_i \neq q_j$.


\phantomsection

\end{document}